\documentclass[letterpaper]{article} 
\usepackage{aaai2026}  
\usepackage{times}  
\usepackage{helvet}  
\usepackage{courier}  
\usepackage[hyphens]{url}  
\usepackage{graphicx} 
\usepackage{natbib}  
\usepackage{caption} 
\usepackage{algorithm}
\usepackage{algorithmic}

\usepackage{newfloat}
\usepackage{listings}
\DeclareCaptionStyle{ruled}{labelfont=normalfont,labelsep=colon,strut=off} 
\floatstyle{ruled}
\newfloat{listing}{tb}{lst}{}
\floatname{listing}{Listing}
\usepackage{enumitem}
\usepackage{amssymb} 
\usepackage{mathtools}
\usepackage{amsthm}
\usepackage{amsmath}
\usepackage{transparent}
\usepackage{color}
\usepackage{xcolor}
\usepackage{colortbl}
\usepackage[makeroom]{cancel}
\usepackage{soul}  
\usepackage{pifont}
\usepackage{rotating} 
\usepackage{float}
\usepackage{booktabs}

\definecolor{gray}{rgb}{0.46,0.46,0.46}
\definecolor{darkergreen}{RGB}{21, 152, 56}
\definecolor{darkerred}{RGB}{220, 35, 120}
\definecolor{darkerblue}{rgb}{0,0.08,0.45} 
\definecolor{royalblue}{RGB}{65,105,225}
\definecolor{lightblue}{RGB}{221,235,247}
\definecolor{gray94}{gray}{.94}
\definecolor{gray90}{gray}{.90}

\newcommand{\gray}[1]{\textcolor{gray}{#1}}

\newcolumntype{g}{>{\columncolor{gray94}}c} 
\newcolumntype{b}{>{\columncolor{lightblue}}c} 
\newcommand{\brow}[1]{\rowcolor{lightblue}{#1}} 
\newcommand{\cmark}{\ding{51}}%
\newcommand{\xmarkg}{\textcolor{gray}{\ding{55}}}%

\title{MergeDNA: Context-aware Genome Modeling with Dynamic Tokenization through Token Merging}

\author{
    Siyuan Li\textsuperscript{\rm 1,2,3},
    Kai Yu\textsuperscript{\rm 2},
    Anna Wang\textsuperscript{\rm 2},
    Zicheng Liu\textsuperscript{\rm 1,2},
    Chang Yu\textsuperscript{\rm 2},
    Jingbo Zhou\textsuperscript{\rm 1,2},\\
    Qirong Yang\textsuperscript{\rm 3}$^*$,
    Yucheng Guo\textsuperscript{\rm 3},
    Xiaoming Zhang\textsuperscript{\rm 3},
    Stan Z. Li\textsuperscript{\rm 2}\thanks{Corresponding authors.}
}
\affiliations{

    \textsuperscript{\rm 1}Zhejiang University, Hangzhou, China\\
    \textsuperscript{\rm 2}AI Lab, Research Center for Industries of the Future, Westlake University, China\\
    \textsuperscript{\rm 3}BioMap Research, Beijing, China\\

}

\usepackage{bibentry}

\begin{document}

\maketitle

\begin{abstract}

Modeling genomic sequences faces two unsolved challenges: the information density varies widely across different regions, while there is no clearly defined minimum vocabulary unit. Relying on either four primitive bases or independently designed DNA tokenizers, existing approaches with naive masked language modeling pre-training often fail to adapt to the varying complexities of genomic sequences.
Leveraging Token Merging techniques, this paper introduces a hierarchical architecture that jointly optimizes a dynamic genomic tokenizer and latent Transformers with context-aware pre-training tasks. As for network structures, the tokenization module automatically chunks adjacent bases into words by stacking multiple layers of the differentiable token merging blocks with local-window constraints, then a Latent Encoder captures the global context of these merged words by full-attention blocks. Symmetrically employing a Latent Decoder and a Local Decoder, MergeDNA learns with two pre-training tasks: Merged Token Reconstruction simultaneously trains the dynamic tokenization module and adaptively filters important tokens, while Adaptive Masked Token Modeling learns to predict these filtered tokens to capture informative contents.
Extensive experiments show that MergeDNA achieves superior performance on three popular DNA benchmarks and several multi-omics tasks with fine-tuning or zero-shot evaluation, outperforming typical tokenization methods and large-scale DNA foundation models.

\end{abstract}


\section{Introduction}
\label{sec:introduction}

Modeling genomic DNA sequences with foundation models~\cite{ji2021dnabert} is an emerging frontier that promises to advance bioinformatics and precision medicine. DNA is often likened to a natural language carrying the ``code of life"~\cite{cooper1981centraldogma}, yet it poses unique modeling challenges far beyond ordinary text.
Firstly, genomic information is distributed unevenly. Only around 2\% of the human genome consists of coding sequences (CDS), densely packed with functional information, whereas the vast majority is non-coding sequence (nCDS) with regulatory or unknown functions, which contains repetitive or less informative content~\cite{science2024EVO}.
Secondly, unlike natural languages with semantic words~\cite{EMNLP2018SentencePiece}, DNA has no inherent word boundaries or predefined vocabulary units~\cite{iclr2024dnabert2}. The meaningful ``units" of DNA vary by context: a biologically relevant motif might be 3 bases (as a codon) \cite{Liu2025LifeCode} or 6–10 bases (a transcription factor binding site), or even longer sequences \cite{dalla2023nucleotide}. This makes fixed tokenization schemes inadequate \cite{nips2024MXDNA}.
Third, DNA sequences are extremely long~\cite{nguyen2024hyenadna}, often spanning tens of thousands to millions of bases, requiring models that can capture both short-range motifs and long-range dependencies efficiently. And naive pre-training objectives~\cite{Radford2018GPT1, devlin2019bert} may fail to focus on the truly important parts of these vast sequences.
These factors collectively make DNA fundamentally distinct from human language and call for a new class of sequence modeling architectures.

\begin{figure}[t]
    \centering
    \vspace{-0.5em}
    \includegraphics[width=1.0\linewidth]{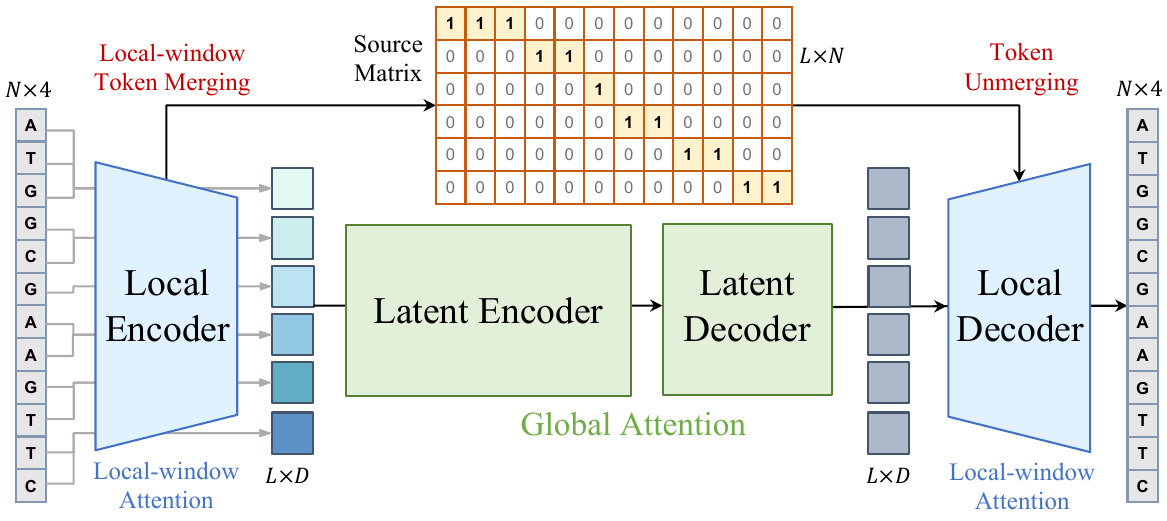}
    \vspace{-1.25em}
    \caption{\textbf{Overview of MergeDNA architecture}. The Local Encoder \& Decoder achieves adaptive DNA tokenization, while the Latent Encoder \& Decoder learn contextual information with informative token masked modeling.
    }
    \label{fig:mergedna_intro}
    \vspace{-1.75em}
\end{figure}

Recent studies have explored various facets of DNA foundation modeling. Long-sequence architectures such as linear-time state-space models (SSMs)~\cite{nguyen2024hyenadna, schiff2024caduceus}, hierarchical Transformers~\cite{NC2024MegaDNA}, and hybrid networks~\cite{science2024EVO, Ma2025HybriDNA} improve context length scalability with efficiency. Meanwhile, DNA tokenization strategies range from base-level encodings to k-mers~\cite{ji2021dnabert, Wu2025GENERator} and learned vocabularies via BPE~\cite{iclr2024dnabert2} or vector quantization~\cite{2017VQ-VAE, icml2024VQDNA}. Pre-training objectives also vary, including masked modeling~\cite{ji2021dnabert}, autoregressive loss~\cite{zhang2023dnagpt}, and advanced masking~\cite{ECAI2023GeneMask}.
However, most works optimize these dimensions in isolation and lack a unified mechanism to address all three DNA modeling challenges. For example, the latest long-range models~\cite{broxiv2025EVO2} that still operate on single-base tokens may waste capacity on repetitive intergenic regions, while a learned tokenizer without a matching long-context encoder could miss global dependencies~\cite{nips2024MXDNA}. In this work, we argue that effective genome-scale modeling requires two core capabilities: (i) a context-sensitive tokenizer that learns to segment DNA into variable-length units based on local structure and semantics, and (ii) adaptive pre-training objectives that prioritize information-dense regions for representation learning.
We try to address these jointly by leveraging token merging techniques~\cite{iclr2022ToMe, nips2024DTEM} for end-to-end learnable token granularity and contextual abstraction.

This work presents \textbf{MergeDNA}, a context-aware genome modeling framework that dynamically adapts tokenization and pre-training to genomic context, as shown in Figure~\ref{fig:mergedna_intro}. The core idea of MergeDNA is a hierarchical autoencoder-style Transformer that learns to compress and reconstruct DNA sequences with a differentiable tokenizer and a long-range context model.
Specifically, we design a Local Encoder composed of stacked local-window attention blocks with differentiable token merging, enabling the model to chunk adjacent bases into variable-length tokens based on local similarity. These merged tokens are then processed by a global-context Latent Encoder using full attention. On the decoder side, a symmetric Latent Decoder and Local Decoder reconstruct the input sequence. Two pre-training objectives jointly supervise the model: (i) Merged Token Reconstruction trains the tokenizer and encoder to preserve key information while filtering redundancies; and (ii) Adaptive Masked Token Modeling selectively masks and predicts important tokens identified through token merging, encouraging context-aware learning of functionally relevant patterns. Together, these components form a unified and scalable genome modeling pipeline that adapts both token resolution and attention allocation based on input complexity.

Our contributions are summarized as follows:
\begin{itemize}
    \item 
    \textbf{Unified Architectural Design:} We propose a novel hierarchical framework that tightly integrates a learnable DNA tokenizer with long-range sequence modeling. Leveraging differentiable token merging within local attention blocks, the Local Encoder captures irregular genomic patterns and determines where to merge as words.
    \item 
    \textbf{Adaptive Context Modeling:} We propose context-aware pre-training tasks that adapt to varying information density in genomic sequences. Using token merging to select informative positions, the proposed Merged Token Reconstruction and Adaptive Masked Token Modeling allow the model to capture both local motif-level information and global long-range dependencies.
    \item 
    \textbf{Strong Empirical Results:} MergeDNA achieves competitive performance across three major DNA benchmarks and shows excellent generalization to several RNA and protein downstream tasks, outperforming prior methods of DNA tokenization and foundation models in both short- and long-context settings.
\end{itemize}

\section{Related Work}
\label{sec:related}

\paragraph{DNA Foundation Models.}

Adapting sequence modeling architectures to genomics has demonstrated impressive transfer capacities to genomic applications, where a family of DNA foundation models~\cite{ji2021dnabert, benegas2023GPN} has merged with four lines of research. \textbf{(a) Long sequence modeling} is the most crucial technique for long DNA sequences. State-space models (SSMs) like HyenaDNA~\cite{nguyen2024hyenadna, thoutam2024MSAMamba} and Caduceus~\cite{schiff2024caduceus} deliver linear complexity, while hierarchical attention \cite{NC2024MegaDNA} or hybrid SSM-attention designs \cite{broxiv2025EVO2, Ma2025HybriDNA} capture both motifs and chromosome-level structure with moderate memory footprints.
\textbf{(b) DNA tokenization} remains discussion with byte-level~\cite{science2024EVO}, k-mers~\cite{dalla2023nucleotide}, or learnable vocabularies~\cite{ji2023genalm, nips2024MXDNA}.
\textbf{(c) Pre-training objectives} can be the BERT-style~\cite{devlin2019bert, iclr2024dnabert2, li2025generanno} or auto-regressive-like masked token modeling~\cite{zhang2023dnagpt, zhu2024CDGPT} for the encoder or decoder architectures, where some loss reweighing~\cite{broxiv2025EVO2} or tailored masking curricula \cite{ECAI2023GeneMask, ECAI2024CMGEMS} could be further beneficial. Only minor methods utilize contrastive learning~\cite{zhou2025dnaberts} or cross-modality alignment tasks~\cite{Liu2025LifeCode} to integrate multi-omic cues.
\textbf{(d) Domains of pre-training and applications} are usually bound. While most models are pre-trained on the human reference~\cite{nguyen2024hyenadna} or multiple species corpora~\cite{iclr2024dnabert2}, specialized datasets confer niche expertise, \textit{e.g.}, prokaryotic~\cite{science2024EVO}, plant genomic domains~\cite{mendoza2023AgroNT, zhai2025PlantCaduceus}, and metagenomes~\cite{zhou2025genomeocean}. Extending beyond monomodal DNA, multi-omics models aim to simulate the central dogma~\cite{cooper1981centraldogma} within a unified architecture~\cite{yang2024CREformer} with a gene-to-expression pipeline~\cite{nm2021enformer, icml2025space} or the genome-to-protein pipeline~\cite{Song2024AIDO, Liu2025LifeCode}, leveraging shared structure across DNA, RNA, protein, and epigenome.

\paragraph{Byte-level Architectures.}
In NLP, early subword approaches like BPE~\cite{Sennrich2015BPE} and SentencePiece~\cite{EMNLP2018SentencePiece} remain the default module in LLMs, and dynamic schemes like Dynamic Pooling~\cite{ACL2023DynamicPool, EMNLP2024DynVoc} partially relax fixed vocabularies, yet they still require external pre-processing. Leveraging SSMs for linear-time attentions~\cite{Gu2023Mamba}, MegaByte~\cite{NIPS2023MEGABYTE}, and MambaByte~\cite{COLM2024Mambabyte, NIPS2024SpaceByte} demonstrated that multi-scale or SSM-based architectures without the subword tokenizer can model million-byte inputs end-to-end with great scale abilities~\cite{Ge2025ByteScale} on text and other modalities~\cite{wu2024bGPT}. More recently, BLT~\cite{Pagnoni2024BLT} introduces learned chunking with entropy-balanced patches, and HNet~\cite{hwang2025hnet} designs differentiable segmentation with jointly optimization. Similarly, DNA models with raw nucleotides as input are also byte-level architectures~\cite{science2024EVO}. Meanwhile, classical tokenization strategies, such as BPE~\cite{iclr2024dnabert2} and k-mers~\cite{dalla2023nucleotide}, as well as learnable dictionaries~\cite{icml2024VQDNA}, have also been explored.


\section{Methodology}
\label{sec:method}

\subsection{Preliminary}
\label{sec:preliminary}
A DNA sequence can be seen as a string in the nucleotide alphabet $\mathcal{D} =\{\texttt{A},\texttt{T},\texttt{C},\texttt{G}\}$. We denote a sequence of length $N$ as $X = (x_1,x_2,\dots,x_N)\in\mathcal{D}^N$, where each $x_i\in\mathcal{D}$. A DNA tokenizer $\mathcal{T}: \mathcal{D}^N \to \mathcal{V}^L$, segments $X$ into a sequence of $L$ tokens $Z_L=(z_1,\dots,z_L)$ and maps to a vocabulary $\mathcal{V}$ with $N \ge L$. Given DNA sequences with a causal mask $M\in \{0,1\}^{N}$, a model $f_{\theta}$ with Attention blocks can be trained with an objective of masked token modeling (MTM):
\begin{equation}
    \mathcal{L}_{MLM}(\theta) = -\frac{1}{L}\sum_{i=1}^L \log P(x_i \mid X * M;\,\theta),
    \label{eq:mlm}
\end{equation}
which encourages $f_{\theta}$ to infer each masked token $x_i$ from its surrounding context to model the DNA context.

\subsection{Architectural Overview}
\label{sec:arch}
Adopting an autoencoder style, MergeDNA consists of four main components, which merge the fixed tokenizer and the sequence model into a hierarchical network in Figure~\ref{fig:mergedna_intro}.

\paragraph{Local Encoder for Tokenization.}
The Local Encoder $\mathcal{E}{\phi}$ serves as a learnable DNA tokenizer with local contexts, producing a tokenized sequence $Z_L\in\mathbb{R}^{L\times D}$ in the embedding dimension of $D$ with a binary source matrix $\mathcal{S} \in \{0,1\}^{L\times N}$:
\begin{equation}
    Z_L,\;\mathcal{S} \;=\; \mathcal{E}_{\phi}(X).
    \label{eq:local_enc}
\end{equation}
Intuitively, $Z_L$ is a context-dependent segmentation of $X$, and each row of $\mathcal{S}$ indicates which original positions in $X$ were merged to form the corresponding token in $Z_L$. 
To adaptively chunk adjacent bases into informative tokens with local context, we implement the Local Encoder as a stack of local-window self-attention layers interleaved with differentiable token merging operations (described in Sec. \ref{sec:mergedna_tok}), where the fixed local window size can also ensure linear-time computational complexity despite the long input.
This learned tokenizer can be trained jointly with the rest of the model, allowing it to optimize token boundaries for the pre-training objective. Rather than using a fixed $k$-mer or requiring a byte-pair scheme, the Local Encoder can allocate shorter tokens (finer granularity) to dense information regions and longer tokens to repetitive regions, thereby addressing the varying information density of genomes.

\paragraph{Latent Context Modeling.}
Based on the tokenized sequence, the Latent Encoder $\mathcal{E}{\psi}$ is the main network for capturing long-range dependencies across the entire input, which can be implemented as a Transformer encoder with full attention (utilizing Flash Attention). As for inference, $\mathcal{E}{\psi}$ produces an output of the same length $L$:
\begin{equation}
    Z'_{L} = \mathcal{E}_{\psi}(Z_L),    
\end{equation}
where $Z'_{L} \in \mathbb{R}^{L\times D}$ are contextually enriched token embeddings. On top of the encoder, we include a lightweight Latent Decoder $\mathcal{E}_{\omega}$, which transforms $Z'_L$ back toward the token space of $\hat{Z_L}$. The Latent Decoder has a symmetric architecture to $\mathcal{E}_{\psi}$, and outputs $\hat{Z}_L = \mathcal{E}_{\omega}(Z'_L)$, where $\hat{Z}_L$ can be seen as a reconstructed version of the Local Encoder’s token sequence, containing the information needed to recover the original input. Together, $\mathcal{E}_{\psi}$ and $\mathcal{E}_{\omega}$ form an autoencoder on the token level. 
This design enables us to apply reconstruction-based training at the token level, providing learning signals to both the tokenizer and the context encoder. We emphasize that the Latent Decoder is used only during pre-training to assist the encoder and tokenizer.
For many downstream tasks, the Latent Decoder may be omitted. This follows the common practice of using an encoder’s learned representations for downstream prediction, while the decoder is specialized for generative reconstruction.

\paragraph{Local Decoder for Reconstruction.}
The final stage is the Local Decoder $\mathcal{E}_{\zeta}$, which maps the Latent Decoder’s output $\hat{Z}_L$ back to the original base space and plays the role of ``detokenizer". We first apply a token unmerging operation $\mathcal{U}(\cdot,\cdot)$ using the source matrix $\mathcal{S}$ to upsampling the $L$-length decoded tokens to length $N$, $\bar{Z}_N = \mathcal{U}(\hat{Z}_L, \mathcal{S})$. $\bar{Z}_N \in \mathbb{R}^{N\times D}$ denotes an unmerged sequence, where each position corresponds to an original base in $X$.
In matrix form, if $\mathcal{S}_{ij} = 1$ indicates the position $i$ covers original position $j$, then $\bar{Z}_N = \mathcal{S}^\top \hat{Z}_L$.
After unmerging, $\mathcal{E}_{\zeta}$ applies a stack of local attention (as the reverse of the Local Encoder) to refine local details and output the reconstructed sequence $\hat{X} = (\hat{x}_1, \dots, \hat{x}_N)$:
\begin{equation}
    \hat{X} \;=\; \mathcal{E}_{\zeta}(\bar{Z}_N).
\end{equation}
The Local Decoder thus completes the autoencoder by learning to fill in base-level information that may have been abstracted away by the Local Encoder. Meanwhile, the Local Encoder can be encouraged to produce merge groupings that are easy to invert, as the source matrix preserves positional information that enables accurate reconstruction.

\paragraph{Training \textit{vs.} Inference.}
During pre-training, we can optimize all modules $\theta = \{\phi,\psi, \omega,\zeta\}$ end-to-end by applying learning objectives of reconstruction and prediction tasks (detailed in Sec.~\ref{sec:pretraining}) between the final output $\hat{X}$ and the original input. 
As for inference, MergeDNA can be truncated or reconfigured depending on the task types, which can function as a typical encoder-only model for representation learning, or as an encoder–decoder model for generative purposes.
For generative tasks or any task requiring output at the nucleotide level,
\textit{e.g.}, sequence reconstruction or base pair prediction,
we can use the entire autoencoder, or fine-tune the Local Decoder for the specific output prediction. For classification or regression tasks at the sample level,
\textit{e.g.}, the goal is to produce a label or embedding for the sequence,
we can discard both decoders and use the Latent Encoder output directly with a fine-tuned head.

\subsection{MergeDNA Tokenization}
\label{sec:mergedna_tok}

\paragraph{Local-window Token Merging.}
At the heart of the Local Encoder is a differentiable token merging mechanism that learns to segment the sequence. We build upon ToMe \cite{iclr2022ToMe}, which progressively fuses similar tokens to reduce sequence length, but adapt it for local, fine-grained chunking. Each Local Encoder layer consists of a standard local self-attention followed by a token merging module. Given the $l$-th layer, supposing the input sequence length is $N_{l-1}$, the merging module will select $r_l$ pairs of tokens within each window to compute the average, reducing the sequence by $r_l$ tokens. We denote this operation as:
\begin{equation}
    \mathcal{S}^{(l)}, Z^{(l)}_{\,N_l} = \text{LocalToMeAttn}^{(l)} \Big(Z^{(l-1)}_{\,N_{l-1}}, \mathcal{S}^{(l-1)}, r_l\Big),
    \label{eq:local_tome}
\end{equation}
where $\mathcal{S}^{(l-1)}$ is the source matrix carried from the previous layer (with $\mathcal{S}^{(0)} = I_{N_0}$ as an identity matrix at input), and $\mathcal{S}^{(l)}$ is updated to reflect the new merges at the $l$-th layer.
In implementation, we compute a similarity score for each pair of tokens in a local window (using a lightweight \textit{grouping} embedding as in DTEM~\cite{nips2024DTEM}). The top-$r_l$ most similar token pairs in each window are selected to merge.
We then perform a \textit{soft merging}: one token (``keeper") absorbs the other (``merger") by adding their representations, or a weighted average, and we mark this in $\mathcal{S}^{(l)}$, where the merger token’s source positions are assigned to the keeper. Tokens not selected for merging pass through unchanged to the next layer. This continuous relaxation of token merging ensures the operation is differentiable, allowing gradients to tune both the token embeddings and the merging criteria.

\begin{figure}[t]
    \centering
    \vspace{-0.5em}
    \includegraphics[width=0.98\linewidth]{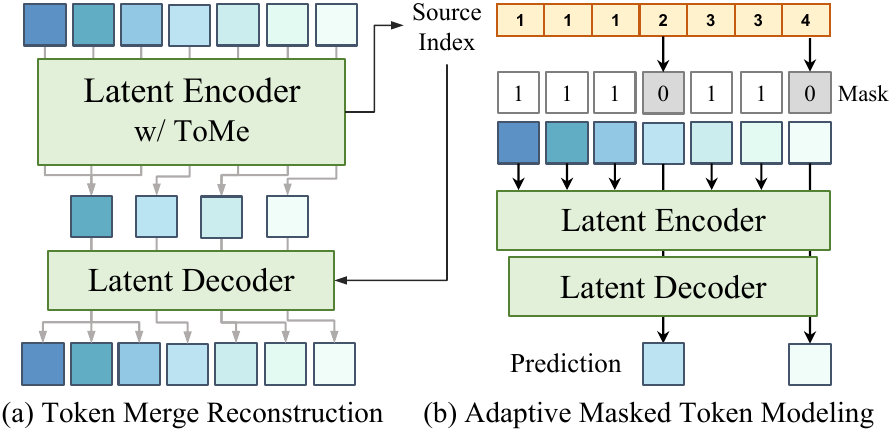}
    \vspace{-0.75em}
    \caption{\textbf{Pre-training of MergeDNA} for (a) Local Encoder \& Decoder and (b) Latent Encoder \& Decoder.
    }
    \label{fig:mergedna_train}
    \vspace{-1.25em}
\end{figure}

\paragraph{Token Unmerging and Reconstruction.}
To optimize the end-to-end tokenization capacity, we introduce a \textit{Merged Token Reconstruction (MTR)} objective $\mathcal{L}_{MTR}$ that forces the network to reconstruct the original sequence from compressed tokens,
which can be computed as the cross-entropy between $\hat{X}$ and $X$:
\begin{equation}
    \mathcal{L}_{MTR}(\theta) = -\frac{1}{N}\sum_{i=1}^N \log P( \hat{X}_i \mid X_i;\; \theta).
\end{equation}
During training, we use a compression ratio sampling strategy, which randomly chooses the number of tokens to retain each iteration. For example, if the average goal is $L \approx \frac{N}{2}$, we might sample $L$ from a Gaussian distribution centered at $\frac{N}{2}$ (with the variance to ensure $L \in [0.4N, 0.6N]$). This strategy exposes $\mathcal{E}{\phi}$ to a wider range of segmentations during training, improving its generalization ability and ensuring that the Local Encoder does not overfit to a particular compression rate.

\begin{table*}[t!]
    \centering
    \vspace{-0.5em}
    \setlength{\tabcolsep}{0.5mm}
\resizebox{1.0\linewidth}{!}{
    \begin{tabular}{l|ccccccccccb}
    \toprule
Method                           & HyenaDNA           & Caduceus-16 & DNABERT    & DNABERT2 & GENA-LM & NT-500M & VQDNA   & MxDNA              & ConvNova & GENERator  & \bf{MergeDNA}    \\
Date                             & \small{NeurIPS'23} & ICML'24     & Bioinfo'21 & ICLR'24  & NAR'23  & NM'24   & ICML'23 & \small{NeurIPS'24} & ICLR'25  & arXiv'25   & \bf{Ours}        \\ \hline
\# Params                        & 6.6M               & 7.9M        & 86M        & 117M     & 113M    & 500M    & 93M     & 100M               & 1.7M     & 1.3B       & 380M             \\
Architecture Type                & byte+SSM           & byte+SSM    & 6-mer+A    & BPE+A    & BPE+A   & 6-mer+A & VQ+A    & DC+A               & byte+CNN & 6-mer+A    & byte+A           \\
Pre-training Task                & AR                 & AR          & BERT       & BERT     & BERT    & BERT    & BERT    & BERT               & BERT     & AR         & \small{MTR+AMTM} \\ \hline
Enhancers (3 tasks)              & 80.88              & 79.96       & 80.14      & 82.81    & 83.22   & 84.56   & 82.37   & 82.79              & 80.90    & \ul{84.87} & \bf{85.11}       \\
Species Classification (2 tasks) & 93.61              & 94.65       & 94.74      & 95.49    & 95.11   & 96.64   & 95.79   & 96.46              & 95.50    & \bf{96.95} & \ul{96.84}       \\
Regulatory Elements (3 tasks)    & 88.89              & 85.97       & 83.42      & 86.33    & 87.89   & 89.05   & 87.62   & \ul{90.57}         & 87.30    & 90.30      & \bf{90.66}       \\ \hline
\bf{Average (8 tasks)}           & 87.07              & 85.89       & 85.02      & 87.30    & 87.94   & 89.26   & 87.69   & 89.12              & 86.95    & \ul{90.71} & \bf{90.87}       \\
    \bottomrule
    \end{tabular}
    }
    \vspace{-0.25em}
    \caption{\textbf{Comparison on Genomic Benchmarks}.  
    Top-1 accuracy (\%) averaged over several similar tasks is reported for popular DNA foundation models with SFT evaluation.  
    The best and the second best results are marked as the \textbf{bold} and \underline{underlined} types.
    }
    \label{tab:genomic_benchmark}
\end{table*}

\begin{table*}[t!]
    \vspace{-0.5em}
    \centering
    \setlength{\tabcolsep}{0.6mm}
\resizebox{1.0\linewidth}{!}{
    \begin{tabular}{l|cccccccccb}
    \toprule
Method                  & HyenaDNA           & Caduceus-PS & DNABERT    & GROVER     & DNABERT2   & NTv2-500M  & MxDNA              & ConvNova   & GENERator  & \bf{MergeDNA} \\
Date                    & \small{NeurIPS'23} & ICML'24     & Bioinfo'21 & bioRxiv'23 & ICLR'24    & NM'24      & \small{NeurIPS'24} & ICLR'25    & arXiv'25   & \bf{Ours}     \\
\# Params (M)           & 6.6M               & 1.9M        & 86M        & 87M        & 117M       & 500M       & 100M               & 1.7M       & 1.2B       & 380M          \\ \hline
H3                      & 78.14              & 80.48       & 77.41      & 76.80      & 79.31      & 78.17      & \ul{82.78}         & 81.50      & 80.60      & \bf{82.95}    \\
H3K4me1                 & 44.52              & 52.83       & 43.83      & 46.10      & 48.34      & 51.64      & 56.15              & \bf{56.60} & 55.30      & \ul{56.24}    \\
H3K4me2                 & 42.68              & 49.88       & 32.38      & 40.30      & 43.02      & 37.24      & 55.59              & \bf{57.45} & 42.40      & \ul{55.67}    \\
H3K4me3                 & 50.41              & 56.72       & 31.49      & 45.80      & 45.43      & 50.30      & 63.68              & \bf{67.15} & 51.20      & \ul{64.10}    \\
H3K9ac                  & 58.50              & 63.27       & 52.55      & 62.60      & 60.04      & 61.05      & 64.78              & \bf{68.10} & 61.20      & \ul{65.01}    \\
H3K14ac                 & 56.71              & 60.84       & 46.51      & 54.80      & 54.49      & 57.22      & 68.27              & \bf{70.71} & 60.50      & \ul{68.51}    \\
H3K36me3                & 59.92              & 61.12       & 50.98      & 56.30      & 57.58      & 60.50      & 67.05              & \ul{68.31} & 65.70      & \bf{68.19}    \\
H3K79me3                & 66.25              & 67.17       & 60.48      & 58.10      & 64.38      & 65.78      & \bf{74.29}         & 72.08      & 67.00      & \ul{74.23}    \\
H4                      & 78.15              & 80.10       & 79.60      & 76.90      & 78.18      & 79.87      & \ul{81.18}         & 81.12      & \bf{81.50} & 81.06         \\
H4ac                    & 54.15              & 59.26       & 41.53      & 53.00      & 51.80      & 55.22      & \bf{67.65}         & 66.10      & 59.20      & \ul{67.26}    \\ \hline
Enhancer                & 53.13              & 55.20       & 79.13      & 51.60      & 52.50      & 54.51      & \bf{79.90}         & 57.60      & 58.00      & \ul{79.84}    \\
Enhancer Types          & 48.16              & 47.17       & 54.73      & 43.30      & 44.32      & 43.36      & \ul{60.50}         & 49.75      & 47.70      & \bf{60.62}    \\
Promoter All            & 95.57              & 96.65       & 97.05      & 92.60      & 96.23      & 96.82      & \ul{97.16}         & 96.82      & 96.20      & \bf{97.40}    \\
Promoter Non-TATA       & 95.86              & 96.31       & 97.02      & 92.50      & 97.17      & \bf{97.45} & 97.24              & 96.76      & 96.20      & \ul{97.35}    \\
Promoter TATA           & 95.88              & 96.21       & 96.22      & 89.10      & \bf{96.99} & 96.53      & 96.01              & 96.34      & 94.80      & \ul{96.70}    \\ \hline
All                     & 94.05              & 92.87       & 97.83      & 91.90      & 93.75      & \ul{98.15} & 98.14              & 96.33      & 97.80      & \bf{98.35}    \\
Accpetor                & 96.98              & 94.21       & 97.81      & 91.20      & 97.49      & 97.99      & 98.01              & 96.23      & \ul{98.10} & \bf{98.67}    \\
Donor                   & 95.27              & 94.69       & 98.43      & 88.80      & 94.33      & \ul{98.50} & 98.10              & 96.62      & 97.80      & \bf{98.93}    \\ \hline
\bf{Average (18 tasks)} & 70.24              & 72.50       & 68.61      & 67.32      & 69.74      & 71.13      & \ul{78.14}         & 76.42      & 72.84      & \bf{78.39}    \\
    \bottomrule
    \end{tabular}
    }
    \caption{\textbf{Comparison on NT Benchmark}. Matthews Correlation Coefficient (MCC) (\%) or F1 score (\%) is reported for sub-tasks with SFT evaluation.  
    The best and the second best results are marked as the \textbf{bold} and \underline{underlined} types.}
    \label{tab:nt_benchmark}
    \vspace{-0.5em}
\end{table*}

\subsection{Adaptive Context Modeling}
\label{sec:pretraining}
As discussed in Sec. \ref{sec:arch}, the Latent Encoder $\mathcal{E}{\psi}$ processes $L$ tokens uniformly during inference. However, genomic sequences often contain long stretches of low-information content (\textit{e.g.}, repetitive DNA) where modeling every token is unnecessary. We aim to help the model find out and focus most informative tokens and design two steps to improve the naive MLM object, as shown in Figure~\ref{fig:mergedna_train}.

\paragraph{Selection and Reconstruction.}
During pre-training, we modify the latent encoder $\mathcal{E}_{\psi}$ to forward an additional round and select a smaller number $K$ of salient tokens. Technically, we replace the standard attention in $\mathcal{E}{\psi}$ with a ToMe-style Attention that merges tokens at the global scale (as opposed to local windows). Formally, we obtain $(Z'_{K}, \mathcal{S}') = \mathcal{E}{\psi}(Z_{L}, \mathcal{S})$, where $Z'_{K} \in \mathbb{R}^{K\times D}$ with $K<L$. 
Note that $\mathcal{S}'$ identifies the $K$ most essential tokens among the $Z_{L}$: the merging algorithm preferentially fuses tokens that appear redundant or less salient, while preserving distinct tokens that carry unique information. We then feed $Z'_K$ into the Latent Decoder to produce $\hat{Z}_L$, but first we upsample it back to length $L$ with the unmerge operation, $\bar{Z}_L = \mathcal{U}(Z'_{K}, \mathcal{S}')$. This distributes each of the $K$ latent tokens back to its original token positions.
Finally, $\hat{Z}_L$ is passed through the Local Decoder to produce $\hat{X}$, and we compute a reconstruction loss as $\mathcal{L}_{MTR}$. We refer to this loss as the latent MTR loss, $\mathcal{L}_{MTR}(\theta \setminus \{\phi\})$, since it trains the latent models to recover context from the selected tokens while the Local Encoder is held fixed ($\phi$ is not updated in this step).
Intuitively, this task forces the latent transformer to not rely on having every token available – it must learn to encode the sequence in such a way that even if nearly $L-K$ tokens worth of information are dropped, the remaining $K$ still capture the essential context to rebuild the sequence. This pushes $\mathcal{E}_{\psi}$ to generate a more compact, salient representation.

\paragraph{Adaptive Masked Token Modeling}
Beyond reconstruction, we also devise a masking strategy to predict the informative tokens. We leverage the latent merging outcome $\mathcal{S}'$ to decide which tokens to mask. The key idea is to assign a higher masking probability to tokens deemed important (those not heavily merged) and a lower probability to tokens that were aggressively merged (with low information).
Given $\mathcal{S}'\in\{0, 1\}^{K\times L}$ from the Latent Encoder, we compute an importance probability for each of the $L$ local tokens. Let $g_i = \sum_{j=1}^L S'\{i,j\}$ be the number of original tokens (out of $L$) that were grouped into the $i$-th latent token. We assign each latent group $i$ a weight inversely proportional to its size, \textit{e.g.}, $w_i = \frac{1}{g_i}$. For each token $j$ that belongs to group $i$, we set $P_L(j) \propto \frac{w_i}{g_i}$, and choose the normalizing constant such that $\sum_{j=1}^L P_L(j)=1$.
This yields a probability vector $P_L \in \mathbb{R}^L$ over the local tokens, where tokens in large merged groups (large $g_i$) receive low probability and tokens in singleton or small groups receive higher probability. We then sample exactly $K$ tokens without replacement according to $P_L$ to mask. Letting $M_L \in \{0,1\}^L$ be the mask indicator, we map this mask back to the input space via the source matrix, $M_N = \mathcal{U}(M_L, \mathcal{S}) \in \{0,1\}^N$. In other words, if a merged token is selected to be masked, all of its constituent base positions in $X$ will be masked out.
Finally, we feed the masked sequence $X*M_{N}$ through the entire network without latent token merging to get an output $\tilde{X}$. We define an Adaptive Masked Token Modeling (AMTM) loss:
\begin{equation}
    \hspace{-0.5em}
    \mathcal{L}_{AMTM}(\theta) = -\frac{1}{K}\sum_{i:\, M_N(i)=1} \log P(\tilde{X}_i \mid X*M_{N}; \theta).
\end{equation}
This is essentially a masked language modeling loss focused on the $K$ high-information tokens (and their base positions), ignoring the easy/redundant tokens.
%
Overall, our full pre-training objectives involve three losses computed in three forward passes, which can be computed as:
\begin{equation}
    \mathcal{L}_{\text{total}} = \mathcal{L}_{MTR}(\theta) + \lambda \mathcal{L}_{MTR}(\theta\setminus\{\phi\}) + \mathcal{L}_{AMTM}(\theta),
\end{equation}
where $\lambda$ denotes a down-weighting factor, we set $\lambda = 0.25$ in practice, which ensures that the model learns to recover dropped information without overweighting the low-information content in its training signal.

\section{Experiments}
\label{sec:exp}

We first evaluate MergeDNA by comparison experiments on DNA benchmarks and multi-omics tasks with a wide range of sequence lengths using the supervised fine-tuning (SFT) or zero-shot evaluation protocols. Then, we empirically analyze the learned vocabularies and the proposed techniques.
All experiments are conducted with PyTorch and NVIDIA A100-80G GPUs for three trials.

\subsection{Experimental Setup}
\label{sec:exp_setting}
\paragraph{Implementations.}
Following the Transformer architecture as LLaMA~\cite{Touvron2023LLaMA}, MergeDNA adopts an embed dimension of $D=1024$ and the local window size of 16. The Local Encoder and Decoder stack 4 and 2 Local ToMeAttention blocks, while the Latent Encoder and Latent Decoder use 20 and 4 Transformer blocks, totaling 380M parameters. Following DNABERT-2 \cite{iclr2024dnabert2}, we pre-train MergeDNA on the \texttt{Multi-Species Genomes} corpus using AdamW optimizer~\cite{iclr2019AdamW} for 100K iterations with a base learning rate of $1\times10^{-4}$ and maximum sequence length of 4096. The hierarchical compression yields a local encoder output length $L$=$\frac{N}{2}$ and a latent encoder length $K$=$\frac{L}{2}$, effectively modeling long-range context with reduced complexity. For downstream tasks, we adhere to the benchmark-specific SFT protocols. On sequence-level classification tasks, we discard both decoders and fine-tune a classification head on the latent encoder’s output. For token-level (base-resolution) tasks, we retain the Local Decoder to recover sequence resolution and fine-tune a new token-level prediction head.
View more details in the Appendix.

\paragraph{Comparison Baselines.}
We compare MergeDNA with state-of-the-art genomics models across four architecture paradigms: (1) sequence modeling architectures with SSMs have HyenaDNA~\cite{nguyen2024hyenadna} and Caduceus~\cite{schiff2024caduceus}, (2) Standard Transformers have DNABERT~\cite{ji2021dnabert}, DNABERT-2~\cite{iclr2024dnabert2}, NTv1/v2~\cite{dalla2023nucleotide}, GROVER~\cite{sanabria2023GROVER}, and GenSLM~\cite{ijhpca2023genslms}), (3) Hybrid models have Evo~\cite{science2024EVO}, Evo2~\cite{broxiv2025EVO2}, and HybriDNA~\cite{Ma2025HybriDNA}), and (4) CNN is ConvNova~\cite{iclr2025convnova}. As for the tokenizer, four popular types are compared in Table \ref{tab:genomic_benchmark}: (1) Byte-level like Evo, (2) k-mer like NTv2, (3) BPE like DNABERT2,  (4) DNA dynamic tokenizer methods have VQDNA~\cite{icml2024VQDNA} and MxDNA~\cite{nips2024MXDNA}. 
All baseline foundation models were pre-trained with standard masked language modeling (BERT) or autoregressive (AR) objectives. As for downstream tasks, typical specialist models are also included.

\begin{table*}[t!]
    \vspace{-1.0em}
    \centering
    \setlength{\tabcolsep}{0.5mm}
\resizebox{1.0\linewidth}{!}{
    \begin{tabular}{l|cccccccccb}
    \toprule
Method                           & HyenaDNA           & Caduceus-PS & DNABERT    & NT-multi   & DNABERT2 & VQDNA      & MxDNA              & ConvNova   & HybriDNA-7B & \bf{MergeDNA} \\
Date                             & \small{NeurIPS'23} & ICML'24     & Bioinfo'21 & NM'24      & ICLR'24  & ICML'24    & \small{NeurIPS'24} & ICLR'25    & arXiv'25    & \bf{Ours}  \\
\# Params (M)                    & 6.6M               & 1.9M        & 86M        & 2.5B       & 117M     & 93M        & 100M               & 1.7M       & 7B          & 380M       \\ \hline
Epigenetic Marks Prediction (10) & 58.94              & 58.39       & 49.08      & 58.06      & 55.98    & 57.95      & 67.29              & \ul{68.91} & 63.05       & \bf{68.82} \\
Human TF Detection (3)           & 61.74              & $-$         & 64.17      & 63.34      & 70.11    & 70.56      & $-$                & $-$        & \bf{72.89}  & \ul{72.24} \\
Mouse TF Detection (3)           & 64.37              & $-$         & 56.43      & 67.02      & 67.99    & 69.80      & $-$                & $-$        & \bf{78.02}  & \ul{73.21} \\
Core Promoter Detection (3)      & 69.22              & $-$         & 71.81      & 71.63      & 70.53    & \ul{73.37} & $-$                & $-$        & 71.37       & \bf{73.41} \\
Promoter Detection (3)           & 80.14              & $-$         & 81.69      & \bf{88.15} & 84.21    & 86.58      & $-$                & $-$        & 85.53       & \ul{87.73} \\
Splice Site Reconstructed (1)    & 77.76              & $-$         & 84.07      & 89.35      & 84.99    & 89.53      & $-$                & $-$        & \bf{90.09}  & \ul{89.95} \\
Virus Covid Classification (1)   & 25.88              & $-$         & 55.50      & 73.04      & 71.02    & \ul{74.32} & $-$                & $-$        & 74.02       & \bf{74.41} \\ \hline
\textbf{Average (24 tasks)}      & 62.58              & 58.39       & 60.53      & 67.23      & 66.43    & 68.51      & 67.29              & 68.91      & \ul{76.42}  & \bf{77.11} \\
    \bottomrule
    \end{tabular}
    }
    \caption{\textbf{Comparison on GUE Benchmark}. Matthews Correlation Coefficient (MCC) (\%) or F1 score (\%) averaged across sub-tasks is reported with SFT evaluation.  
    The best and the second best results are marked as the \textbf{bold} and \underline{underlined} types.}
    \label{tab:gue}
    \vspace{-0.5em}
\end{table*}

\subsection{Comparison Results on Genomic Benchmarks}
\label{sec:dna_comparison}

\paragraph{Genomic Benchmarks.}
We first evaluate on eight representative tasks from the Genomic Benchmark suite~\cite{BMC2023genomicbenchmark}, covering enhancer identification, species classification, and regulatory element prediction. All models are fine-tuned on each task, and we report top-1 accuracy following the GenBench protocol. As shown in Table~\ref{tab:genomic_benchmark}, MergeDNA achieves the highest overall accuracy (90.87\%), outperforming all prior DNA foundation models. Notably, it yields state-of-the-art results on the enhancer tasks (85.11\% \textit{vs} 84.87\% by the second best) and regulatory element tasks, while maintaining competitive performance on species classification (second only to a larger model).
These improvements underscore the advantages of our context-aware tokenizer and hierarchical modeling in understanding generic genomic sequences.

\paragraph{Nucleotide Transformer Benchmarks.}
We further compare on the comprehensive Nucleotide Transformer (NT) benchmark (18 tasks)~\cite{dalla2023nucleotide} as summarized in Table~\ref{tab:nt_benchmark}. This benchmark includes a diverse mix of epigenomic signal classification (various histone marks and enhancer states, measured by MCC or F1) and core promoter/splice site detection tasks.
MergeDNA again attains the best overall performance with an average score of 78.39, slightly surpassing the previous dynamic tokenizer method MxDNA (78.14) and substantially higher than other baselines. In particular, our model consistently ranks at or near the top on most individual tasks (\textit{e.g.}, achieving the highest MCC on 10 out of 18 tasks).

\paragraph{GUE Benchmarks.}
We also evaluate on the Genome Understanding Evaluation (GUE) suite introduced by DNABERT2~\cite{iclr2024dnabert2}, which aggregates 24 short-range subtasks grouped into seven practical genomic applications: Epigenetic Mark Prediction (Yeast), Transcription Factor (TF) binding site detection (Human and Mouse), Promoter and Core Promoter Detection, Splice Site Prediction, and Virus Genomic Classification. We use the Matthews Correlation Coefficient (MCC) or the F1 score, as in prior work, and baseline results are taken from DNABERT-2 or the original papers for consistency. As summarized in Table~\ref{tab:gue}, MergeDNA delivers the highest mean performance (77.11\%), edging out the much larger HybriDNA-7B (76.42\%) and outperforming all other foundation models.
These results highlight that MergeDNA’s dynamic tokenization and dual-context pre-training yield broad improvements across heterogeneous genomic prediction tasks.


\begin{table}[t!]
    \centering
    \setlength{\tabcolsep}{0.5mm}
\resizebox{1.0\linewidth}{!}{
    \begin{tabular}{l|cccccb}
    \toprule
Method    & SpliceAI & DNABERT2 & NT-500M & Caduceus & Evo2-7B   & \bf{MergeDNA} \\
\# Params & 3.5M     & 117M     & 500M    & 7.9M     & 7B        & 380M          \\ \hline
Donor     & 57.4     & 63.5     & 55.7    & 64.2     & \bf{64.5} & \ul{64.4}     \\
Acceptor  & 69.1     & 70.7     & 72.2    & 74.0     & \ul{74.3} & \bf{74.5}     \\
Mean      & 63.2     & 67.1     & 63.9    & 69.1     & \ul{69.2} & \bf{69.8}     \\
    \bottomrule
    \end{tabular}
    }
    \vspace{-0.25em}
    \caption{\textbf{Comparison on Splicing Prediction} on the SpliceAI dataset, where the AUROC score is reported. 
    }
    \label{tab:splicing}
    \vspace{-0.5em}
\end{table}

\begin{table}[t!]
    \centering
    \setlength{\tabcolsep}{0.4mm}
\resizebox{1.0\linewidth}{!}{
    \begin{tabular}{l|cccccb}
    \toprule
Method    & GenSLM-2.5B & NT-2500M & \gray{ESM2-650M} & Evo-7B     & Evo2-7B    & \bf{MergeDNA} \\
\# Params & 2.5B        & 2.5B     & \gray{650M}      & 7B         & 7B         & 380M          \\ \hline
Bacteria  & 24.7        & 9.4      & \gray{51.2}      & \ul{45.30} & \bf{45.85} & 42.72         \\
Human     & 6.9         & 4.7      & \gray{37.5}      & 11.10      & \bf{36.9}  & \ul{20.58}    \\
    \bottomrule
    \end{tabular}
    }
    \vspace{-0.25em}
    \caption{\textbf{Comparison on Protein Fitness Prediction}. Zero-shot SRCC (\%) is reported on DMS datasets.
    }
    \label{tab:protein_dms}
    \vspace{-0.5em}
\end{table}

\subsection{Multi-omics Downstream Tasks}
\label{sec:multi_omics}
We further assess MergeDNA’s generalization to long-range and cross-omics scenarios, including RNA splicing, expression prediction, and protein tasks.

\paragraph{RNA Splicing Site Prediction.}
Pre-mRNA splicing is a crucial step in gene expression, and we evaluate our model on the SpliceAI dataset~\cite{JAGANATHAN2019SpliceAI}, which provides long pre-mRNA sequences labeled with donor and acceptor splice sites. We treat this as a binary sequence classification (site vs. non-site) and report the area under the ROC curve (AUROC). In Table~\ref{tab:splicing}, MergeDNA achieves a mean AUROC of 69.8, substantially outperforming the classic SpliceAI model (63.2) and all prior DNA foundation models.
MergeDNA nearly matches the 7B-parameter Evo2 on donor site prediction and exceeds it on acceptor sites.

\begin{figure*}[t!]
    \centering
    \vspace{-0.5em}
    \includegraphics[width=\linewidth]{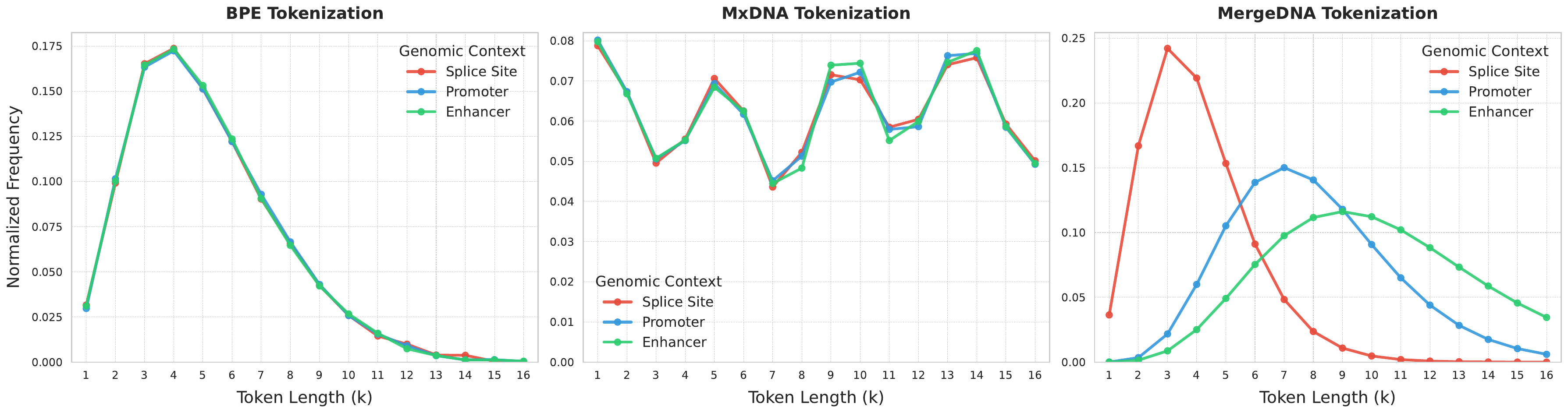}
    \vspace{-1.75em}
    \caption{\textbf{Visualization of Token Length Distributions} for (a) BPE~\cite{iclr2024dnabert2}, (b) MxDNA~\cite{nips2024MXDNA}, and (c) MergeDNA across different genomic contexts. Baseline tokenizers show a static, context-agnostic distribution, while MergeDNA adaptively changes its tokenization strategy based on the sequence type, demonstrating strong context-awareness.}
    \label{fig:tokenization_comparison}
    \vspace{-0.5em}
\end{figure*}

\begin{table*}[t]
    \centering
\resizebox{0.98\linewidth}{!}{
    \begin{tabular}{l|cccccccb}
    \toprule
Method              & DNABERT2 & DNABERT-S & NTv2-500M & HyenaDNA-160K & Caduceus-131K & HybridDNA-131K & Evo2-7B   & \bf{MergeDNA} \\
\# Params           & 117M     & 117M      & 500M      & 12.9M         & 7.7M          & 300M           & 7B        & 380M          \\ \hline
Causal eQTL (AUROC) & 0.72     & 0.73      & 0.72      & 0.71          & 0.68          & \ul{0.74}      & \ul{0.74} & \bf{0.75}     \\
Bulk RNA ($R^2$)    & 0.51     & 0.52      & \ul{0.60} & 0.46          & 0.52          & 0.52           & \ul{0.60} & \bf{0.62}     \\
    \bottomrule
    \end{tabular}
    }
    \vspace{-0.25em}
    \caption{
    \textbf{Comparison on LRB Benchmark} with Causal eQTL Variant Effect Prediction and Bulk RNA Expression Prediction, where AUROC and $R^2$ scores are reported.
    }
    \label{tab:lrb}
    \vspace{-0.5em}
\end{table*}

\paragraph{Long-range Expression Prediction.}
We next consider two challenging expression quantitative trait tasks from the Genomics Long-Range Benchmark (LRB)~\cite{trop2024LRB} that demand modeling of kilobase-scale contexts. (i) Causal eQTL Effect Prediction: given a genomic locus and a candidate variant, predict if the variant alters gene expression (evaluated by AUROC). (ii) Bulk RNA Expression Prediction: predict gene expression levels from the surrounding DNA sequence (evaluated by $R^2$ correlation). As shown in Table~\ref{tab:lrb}, MergeDNA attains new state-of-the-art results on both tasks: an AUROC of 0.75 for eQTL (\textit{vs} 0.74 by the best baseline) and an $R^2$ of 0.62 for bulk expression (\textit{vs} 0.60).

\paragraph{Protein Fitness Prediction.}
Finally, we further evaluate MergeDNA in a strict zero-shot setting on protein fitness prediction tasks using Deep Mutational Scanning (DMS) data~\cite{icml2022Tranception}. Here, models must predict the functional fitness of protein variants (amino acid mutations) directly from the DNA coding sequence, without any fine-tuning on protein data. Table~\ref{tab:protein_dms} reports Spearman’s rank correlation coefficient (SRCC) between predicted and actual fitness on two representative DMS datasets (one bacterial protein and one human protein). A specialized protein language model (ESM2~\cite{lin2022ESM2}, gray in the table) achieves the highest scores as expected. Among DNA-based models, MergeDNA shows strong cross-omics generalization: for the bacterial protein, it obtains 42.7\% SRCC—on par with the 7B Evo model and only slightly behind the multi-omics Evo2 (45.9\%). On the human protein, MergeDNA (20.6\% SRCC) substantially outperforms earlier DNA models like GenSLM (6.9\%) and the original Evo (11.1\%), though it trails Evo2, which leverages direct protein training.

\subsection{Empirical Analysis of Tokenization}
\label{sec:exp_empirical}
To understand how MergeDNA learns to parse genomic sequences, we analyze the vocabularies learned by different tokenization strategies. We compare MergeDNA with two representative baseline tokenizers, \textbf{BPE} and \textbf{MxDNA}, by visualizing the normalized frequency of token lengths (from 1-mer to 16-mer) across different genomic contexts: promoters, enhancers, and splice sites.
%
The BPE tokenizer in Figure~\ref{fig:tokenization_comparison}(a) produces a fixed, long-tailed distribution that peaks around a token length of 6, regardless of the underlying sequence type. Similarly, the vector-quantized MxDNA tokenizer in Figure~\ref{fig:tokenization_comparison}(b) yields a relatively uniform token length distribution that also shows minimal variation across different genomic contexts. This context-agnostic behavior limits their ability to adaptively capture functionally relevant motifs of varying lengths.
%
In sharp contrast, MergeDNA's local encoder, as shown in Figure~\ref{fig:tokenization_comparison}(c), demonstrates strong context-awareness. It learns to produce different token length distributions tailored to the biological properties of the input sequence. 
For splice sites, which are characterized by short, conserved motifs, the distribution peaks at a shorter token length like $k=4$.
For longer and more complex regulatory regions like promoters and enhancers, the distributions shift towards longer tokens (peaking at $k=7$ and $k=9$, respectively). This data-driven tokenization allows MergeDNA to dynamically capture genomic motifs at their relevant biological scales, providing a more expressive input representation for the downstream encoder and contributing to its superior performance.

\begin{table}[htb]
    \centering
    \setlength{\tabcolsep}{0.4mm}
\resizebox{1.0\linewidth}{!}{
    \begin{tabular}{ccc|c|c}
    \toprule
Tokenizer      & Latent Enc. & Local Dec. & Pre-training                                                                                                & Acc.  \\ \hline
Byte           & 24 layers   & 2 layers   & $\mathcal{L}_{MTM}$                                                                                         & 89.30       \\
Local Enc. (4) & 20 layers   & 2 layers   & $\mathcal{L}_{MTR}^{\theta}$ + $\mathcal{L}_{MTM}$                                                            & +0.39       \\
Local Enc. (4) & 20 layers   & 2 layers   & $\mathcal{L}_{MTR}^{\theta}$ + $\mathcal{L}_{MTR}^{\theta \setminus \{\phi\}}$ +$\mathcal{L}_{AMTM}$      & +1.03       \\
\brow Local Enc. (4) & 20 layers   & 2 layers   & $\mathcal{L}_{MTR}^{\theta}$ + $\lambda \mathcal{L}_{MTR}^{\theta \setminus \{\phi\}}$ +$\mathcal{L}_{AMTM}$ & \bf{+1.57}  \\
Local Enc. (2) & 20 layers   & 4 layers   & $\mathcal{L}_{MTR}^{\theta}$ + $\lambda \mathcal{L}_{MTR}^{\theta \setminus \{\phi\}}$ +$\mathcal{L}_{AMTM}$ & +1.21      \\
    \bottomrule
    \end{tabular}
    }
    \vspace{-0.25em}
    \caption{\textbf{Ablation Study} of MergeDNA and pre-training tasks with DNA, protein, and central dogma (CD) tasks. Note that the blue background denotes the selected setups.
    }
    \label{tab:ablation}
    \vspace{-1.5em}
\end{table}

\subsection{Ablation Study}
We conduct ablation experiments on the Genomic Benchmark tasks to quantify the contribution of each component in MergeDNA. Table~\ref{tab:ablation} summarizes the average top-1 accuracy on the 8-task Genomic Benchmark under various configurations.
First, we examine the impact of the hierarchical architecture. Replacing the first 4 Transformer layers with our Local Encoder (merging tokens in windows) improves performance by +0.39 with the same parameter budget, confirming the benefit of local token merging.
Next, we ablate the pre-training objectives. Training with only the naive masked token modeling (MTM) objective results in suboptimal performance; adding the Merged Token Reconstruction ($\mathcal{L}{MTR}$) objectives targeting the tokenizer’s output provides a large gain (+1.03), and further introducing the Adaptive MTM on filtered tokens pushes the improvement to +1.57 over baseline. We also find that scaling down the loss weight $\lambda$ for the latent $\mathcal{L}{MTR}$ (\textit{i.e.}, not directly updating tokenizer parameters) to 0.25 is crucial for better generalization, yielding the best result.
Finally, we vary the depth of the Local Encoder: using only 2 local merging layers (with correspondingly more latent decoder layers) degrades performance, indicating that a deeper tokenizer (4 merging blocks) is important for capturing rich subword representations.

\section{Conclusions and Limitations}
\label{sec:conclusion}

\paragraph{Conclusions.}
We introduce MergeDNA, a context-aware DNA foundation model that addresses fundamental challenges in genome modeling: heterogeneous information density, ambiguous sequence tokenization, and long-range dependencies. MergeDNA unifies a differentiable local tokenizer and a global latent Transformer through a hierarchical architecture and two complementary pre-training tasks, \textit{i.e.}, Merged Token Reconstruction and Adaptive Masked Token Modeling.
These innovations enable the model to dynamically adjust token granularity and focus on salient regions across diverse genomic contexts.
Extensive experiments on three standard DNA benchmarks and several multi-omics tasks demonstrate that MergeDNA achieves state-of-the-art performance with strong generalization across species and modalities, offering a scalable and principled approach to genome-scale representation learning.

\paragraph{Limitations and Future Work.}
Our study is subject to several practical limitations that also suggest promising avenues for future research.
(1) \textit{Genome-scale contexts}. MergeDNA is pre-trained on sequences up to 4k bases and evaluated on tasks spanning up to tens of kilobases via windowed processing and aggregation. While this setup already covers many regulatory elements, fully genome-scale modeling at chromosome- or genome-wide resolution remains beyond the current scope. Extending MergeDNA to 100k–1M context lengths will likely require additional hierarchical stages, more aggressive yet structured token compression, and improved memory management.
(2) \textit{Interpretability of dynamic tokenization}. Currently, we do not yet provide a systematic biological interpretation of the resulting merged tokens. In particular, it remains an open question how frequently merged tokens align with canonical motifs such as splice sites, transcription factor binding sites, or promoter elements, and how tokenization behaves in repetitive or low-complexity regions. Future work could integrate motif discovery tools, attribution methods, and sequence perturbation analyses to connect token merging decisions more directly.
(3) \textit{Connections from DNA tokenizers to tokenizer-free models}.
Conceptually, MergeDNA can be viewed as a dynamic tokenizer tightly coupled with a hierarchical encoder–decoder, positioned between static k-mer/BPE-based DNA tokenizers and fully tokenizer-free byte-level architectures. In this work, we do not systematically explore hybrid designs that combine MergeDNA with recent tokenizer-free long-sequence models, nor do we examine using MergeDNA as a front-end for such encoders. Designing architectures that treat merged tokens as adaptive ``patches'' for downstream byte-level models, or that incorporate differentiable segmentation ideas from tokenizer-free sequence modeling, may further improve the trade-off between flexibility, scalability, and modeling fidelity.
(4) \textit{Multi-omics integration}.
Currently, MergeDNA is pre-trained on the Multi-Species Genomes corpus and evaluated on a representative but still limited set of genomic and multi-omics benchmarks. Although the model generalizes to several downstream DNA, RNA, and protein tasks, broader evaluations in single-cell, epigenomic, metagenomic, and clinical variant interpretation settings are necessary to characterize its strengths and limitations thoroughly.

\section*{Acknowledgments}
This work was supported by the National Natural Science Foundation of China (Project No. 624B2115, 623B2086, and U21A20427), the Science \& Technology Innovation 2030 Major Program (Project No. 2021ZD0150100), the Center of Synthetic Biology and Integrated Bioengineering at Westlake University (Project No. WU2022A009), and the Westlake University Industries of the Future Research Program (Project No. WU2023C019).
This work was done when Siyuan Li interned at BioMap Research. We thank the GPU support from BioMap Research and the AI station of Westlake University.




\bibliography{reference}


\clearpage
\renewcommand\thefigure{A\arabic{figure}}
\renewcommand\thetable{A\arabic{table}}
\setcounter{table}{0}
\setcounter{figure}{0}

\appendix


\section{Implementation Details}
\label{app:implement}
\subsection{Pre-training Settings}
Following \cite{dalla2023nucleotide, iclr2024dnabert2}, we utilize a pure DNA dataset for pre-training, \texttt{Multi-species Genomes} \footnote{Multi-species Genomes are originally provided in \url{https://huggingface.co/datasets/InstaDeepAI/multi_species_genomes}, which is further extended by DNABERT2 in \url{https://github.com/MAGICS-LAB/DNABERT_2}}, with reference sequences (RefSeq) of multiple species to ensure generalization abilities of multiple domains from the National Center for Biotechnology Information (NCBI) database at \url{https://www.ncbi.nlm.nih.gov}.
We pre-train the MergeDNA model (including the latent decoder and the local decoder) by AdamW optimizer~\cite{iclr2019AdamW} for 100,000 iterations (randomly sampled datasets) with a basic learning rate of $1\times 10^{-4}$ and a batch size of 256. With the Ubuntu workstation, the model is pre-trained by 8 Nvidia A100-80G GPUs with a per-GPU batch size of 8 and a gradient accumulation time of 16 for nearly 5 days.
The experiments are implemented on a host machine in Ubuntu (the kernel version of 5.4.0), which is equipped with 8 A100-80G GPUs, 128 CPU cores, and a total memory of 1024 GB.

\begin{table}[ht]
    \setlength{\tabcolsep}{1.0mm}
    \centering
\resizebox{1.0\linewidth}{!}{
\begin{tabular}{l|cccc}
    \toprule
Configuration       & Local Enc. & Latent Enc. & Latent Dec. & Local Dec. \\ \hline
Embedding dim       & \multicolumn{4}{c}{1024}                            \\
Block number        & 4          & 20          & 4           & 2          \\
Block type          & Local-Attn & Attention   & Attention   & Local-Attn \\
Permanent           & \cmark     & \cmark      & \xmarkg     & \xmarkg    \\
Parameters          & 51M        & 253M        & 51M         & 25M        \\ \hline
Optimizer           & \multicolumn{4}{c}{AdamW}                           \\
$(\beta_1,\beta_2)$ & \multicolumn{4}{c}{$(0.9,0.95)$}                    \\
Training iterations & \multicolumn{4}{c}{100,000}                         \\
Weight decay        & \multicolumn{4}{c}{$1\times 10^{-8}$}               \\
Base learning rate  & \multicolumn{4}{c}{$1\times 10^{-4}$}               \\
Batch size          & \multicolumn{4}{c}{256}                             \\
LR scheduler        & \multicolumn{4}{c}{Cosine Annealing}                \\
Warmup iterations   & \multicolumn{4}{c}{10,000}                          \\
Gradient clipping   & \multicolumn{4}{c}{1.0}                             \\
    \bottomrule
    \end{tabular}
    }
    \vspace{-0.5em}
    \caption{
    Configuration of the network architecture and pre-training for MergeDNA. The Local-Attn or Attention blocks denote the local-window or self-attention block with Flash-Attention implementation.
    }
    \label{tab:app_config}
\end{table}

\subsection{Evaluation Setups}
\label{app:impl_evaluation}
In most cases, we apply Supervised Fine-tuning (SFT) to evaluate the transfer capacity of pre-trained models to genomic and multi-omics downstream tasks. Following \citep{nguyen2024hyenadna, iclr2024dnabert2}, adding the decoder head (\textit{e.g.}, an MLP head) to a specific downstream task, the linear attention (RNN) or self-attention blocks in the pre-trained encoder models are frozen, while the Low-Rank Adaptation (LoRA) strategy~\citep{Hu2021LoRA} is employed to parameter-efficiently fine-tune the models by AdamW optimizer with a batch size of 32. For each task, if the benchmark and models have provided hyper-parameters, we follow the official settings, or we choose the best combinations of the basic learning rate \{$1\times 10^{-5}$, $5\times 10^{-5}$, $1\times 10^{-4}$\}, the weight decay \{0, 0.01\}, the LoRA rank \{4, As for the Protein Fitness Prediction task, we conduct zero-shot evaluation by learning a linear regression model upon the embedding of the Latent Encoder. GenBench~\citep{liu2024genbench}. 
As for the Protein Fitness Prediction task, we conduct zero-shot evaluation by learning a linear regression model upon the embedding of the Latent Encoder.
Overall, we report the averaged results over three runs with the optimal settings.

\begin{table*}[t]
    \centering
    \setlength{\tabcolsep}{0.5mm}
\resizebox{1.0\linewidth}{!}{
    \begin{tabular}{l|ccccccccccb}
    \toprule
Method                  & HyenaDNA           & Caduceus-16 & DNABERT    & DNABERT2 & GENA-LM    & NT-500M & VQDNA   & MxDNA              & ConvNova & GENERator  & \bf{MergeDNA} \\
Date                    & \small{NeurIPS'23} & ICML'24     & Bioinfo'21 & ICLR'24  & NAR'23     & NM'24   & ICML'23 & \small{NeurIPS'24} & ICLR'25  & arXiv'25   & \bf{Ours}     \\
\# Params               & 6.6M               & 7.9M        & 86M        & 117M     & 113M       & 500M    & 93M     & 100M               & 1.7M     & 1.3B       & 380M          \\ \hline
Mouse Enhancers         & 79.34              & 81.63       & 80.99      & 81.82    & 82.97      & 85.12   & 81.06   & 80.57              & 78.40    & \bf{87.10} & \ul{85.62}    \\
Human Enhancers Cohn    & 72.96              & 73.76       & 70.23      & 75.87    & 75.63      & 76.12   & 75.63   & 74.67              & 74.30    & \ul{76.30} & \bf{76.54}    \\
Human Enhancers Ensembl & 90.33              & 84.48       & 89.19      & 90.75    & 91.07      & 92.44   & 90.41   & \ul{93.13}         & 90.00    & 91.20      & \bf{93.18}    \\ \hline
Coding vs Intergenomic  & 90.97              & 93.72       & 93.64      & 93.58    & 93.24      & 95.76   & 94.35   & 95.28              & 94.30    & \ul{95.90} & \bf{96.02}    \\
Human vs Worm           & 96.24              & 95.57       & 95.84      & 97.39    & 96.98      & 97.51   & 97.23   & 97.64              & 96.70    & \bf{98.00} & \ul{97.65}    \\ \hline
Human Regulatory        & 93.08              & 87.30       & 88.16      & 87.94    & 88.10      & 93.79   & 90.92   & \bf{94.11}         & 87.30    & 92.80      & \ul{93.49}    \\
Human OCR Ensembl       & 79.14              & 81.76       & 74.96      & 75.82    & 78.98      & 80.42   & 76.58   & 81.05              & 79.30    & \bf{82.30} & \ul{82.16}    \\
Human NonTATA Promoters & 94.45              & 88.85       & 87.13      & 95.24    & \bf{96.60} & 92.95   & 95.37   & 96.56              & 95.30    & 95.80      & \ul{96.34}    \\
    \bottomrule
    \end{tabular}
    }
    \vspace{-0.25em}
    \caption{\textbf{Full Results on Genomic Benchmarks}. Top-1 accuracy (\%) averaged across three trials is reported for the latest DNA foundation models, where the best and the second best results are marked as the \textbf{bold} and \underline{underlined} types.
    }
    \label{tab:app_genomic_benchmark}
    \vspace{-0.5em}
\end{table*}

\begin{table*}[t]
    \centering
    \setlength{\tabcolsep}{0.8mm}
\resizebox{1.0\linewidth}{!}{
    \begin{tabular}{l|cccccccccb}
    \toprule
Method                     & HyenaDNA           & Caduceus-PS & DNABERT    & NT-2500M-multi & DNABERT2 & VQDNA      & MxDNA              & ConvNova   & HybriDNA-7B & \bf{MergeDNA} \\
Date                       & \small{NeurIPS'23} & ICML'24     & Bioinfo'21 & NM'23          & ICLR'24  & ICML'24    & \small{NeurIPS'24} & ICLR'25    & arXiv'25    & \bf{Ours}     \\
\# Params                  & 6.6M               & 1.9M        & 86M        & 2.5B           & 117M     & 93M        & 100M               & 1.7M       & 7B          & 380M          \\ \hline
Human TF-0                 & 64.47              & $-$         & 66.84      & 66.64          & 71.99    & \bf{72.48} & $-$                & $-$        & 70.00       & \ul{72.36}    \\
Human TF-1                 & 70.74              & $-$         & 70.14      & 70.28          & 76.0     & \ul{76.43} & $-$                & $-$        & 74.47       & \bf{76.50}    \\
Human TF-2                 & 60.44              & $-$         & 61.03      & 58.72          & 66.52    & 66.80      & $-$                & $-$        & \ul{70.42}  & \bf{70.48}    \\
Human TF-3                 & 39.78              & $-$         & 51.89      & 51.65          & 58.54    & 58.92      & $-$                & $-$        & \bf{64.52}  & \ul{61.12}    \\
Human TF-4                 & 73.27              & $-$         & 70.97      & 69.43          & 77.43    & 78.10      & $-$                & $-$        & \bf{85.03}  & \ul{80.76}    \\ \hline
Mouse TF-0                 & 56.25              & $-$         & 44.42      & 63.31          & 56.76    & 58.34      & $-$                & $-$        & \bf{71.68}  & \ul{68.23}    \\
Mouse TF-1                 & 80.46              & $-$         & 78.94      & 83.76          & 84.77    & 85.81      & $-$                & $-$        & \bf{87.75}  & \ul{86.69}    \\
Mouse TF-2                 & 78.14              & $-$         & 71.44      & 71.52          & 79.32    & 80.39      & $-$                & $-$        & \bf{86.59}  & \ul{82.85}    \\
Mouse TF-3                 & 60.83              & $-$         & 44.89      & 69.44          & 66.47    & 69.72      & $-$                & $-$        & \bf{87.62}  & \ul{73.46}    \\
Mouse TF-4                 & 46.25              & $-$         & 42.48      & 47.07          & 52.66    & 54.73      & $-$                & $-$        & \bf{56.47}  & \ul{54.82}    \\ \hline
Core Promoter (all)        & 66.18              & $-$         & 68.90      & 70.33          & 69.37    & \bf{71.02} & $-$                & $-$        & 66.50       & \ul{70.78}    \\
Core Promoter (no TATA)    & 67.41              & $-$         & 70.47      & \bf{71.58}     & 68.04    & 70.58      & $-$                & $-$        & 70.66       & \bf{70.91}    \\
Core Promoter (TATA)       & 74.07              & $-$         & 76.06      & 72.97          & 74.17    & \ul{78.50} & $-$                & $-$        & 76.94       & \bf{78.54}    \\ \hline
Promoter (all)             & 83.04              & $-$         & 90.48      & \ul{91.01}     & 86.77    & 90.75      & $-$                & $-$        & 88.28       & \bf{91.02}    \\
Promoter (no TATA)         & 91.03              & $-$         & 93.05      & 94.00          & 94.27    & 94.40      & $-$                & $-$        & \ul{94.73}  & \bf{94.90}    \\
Promoter (TATA)            & 66.36              & $-$         & 61.56      & \bf{79.43}     & 71.59    & 74.52      & $-$                & $-$        & 73.59       & \ul{77.27}    \\ \hline
Splice Reconstructed       & 77.76              & $-$         & 84.07      & 89.35          & 84.99    & 89.50      & $-$                & $-$        & \bf{90.09}  & \ul{89.95}    \\ \hline
H3                         & 78.14              & 77.90       & 73.10      & 78.77          & 78.27    & 79.21      & \bf{82.14}         & 81.50      & $-$         & \ul{81.63}    \\
H3K14ac                    & 56.71              & 54.10       & 40.06      & 56.20          & 52.57    & 54.46      & 68.29              & \bf{70.71} & $-$         & \ul{69.09}    \\
H3K36me3                   & 59.92              & 60.90       & 47.25      & 61.99          & 56.88    & 61.75      & 65.46              & \bf{68.31} & $-$         & \ul{68.24}    \\
H3K4me1                    & 44.52              & 48.80       & 41.44      & 55.30          & 50.52    & 53.28      & 54.97              & \ul{56.60} & $-$         & \bf{57.10}    \\
H3K4me2                    & 42.68              & 38.80       & 32.27      & 36.49          & 31.13    & 34.05      & 55.30              & \bf{57.45} & $-$         & \ul{55.87}    \\
H3K4me3                    & 50.41              & 44.00       & 27.81      & 40.34          & 36.27    & 39.10      & 63.80              & \bf{67.15} & $-$         & \ul{66.84}    \\
H3K79me3                   & 66.25              & 67.60       & 61.17      & 64.70          & 67.39    & 68.47      & \ul{73.74}         & 72.08      & $-$         & \bf{73.80}    \\
H3K9ac                     & 58.50              & 60.40       & 51.22      & 56.01          & 55.63    & 56.63      & 63.15              & \ul{68.10} & $-$         & \bf{68.36}    \\
H4                         & 78.15              & 78.90       & 79.26      & 81.67          & 80.71    & \ul{81.84} & 80.89              & 81.12      & $-$         & \bf{82.06}    \\
H4ac                       & 54.15              & 52.50       & 37.24      & 49.13          & 50.43    & 50.69      & 65.14              & \bf{66.10} & $-$         & \ul{65.22}    \\ \hline
Virus Covid Classification & 25.88              & $-$         & 55.50      & 73.04          & 71.02    & \ul{74.32} & $-$                & $-$        & 74.02       & \bf{74.41}    \\
    \bottomrule
    \end{tabular}
    }
    \vspace{-0.25em}
    \caption{\textbf{Full Results on GUE benchmark}. MCC (\%) or F1 (\%) is reported for Epigenetic Marks Prediction, Human Transcription Factor (TF) Prediction, Mouse Transcription Factor Prediction, Core Promoter Detection, Promoter Detection, Splice Site Reconstructed, and Covid Variants Classification (Virus Covid). The best and the second best results are marked as the \textbf{bold} and \underline{underlined} types.
    }
    \label{tab:app_gue}
    \vspace{-0.5em}
\end{table*}

\section{Downstream Task Benchmarks}
\label{app:downstream}
\subsection{DNA Tasks with Genomics Benchmark}
As proposed by \citep{BMC2023genomicbenchmark}, three groups of basic genomic tasks are collected as balanced binary classification with top-1 accuracy in the Genomics Benchmark. Regarding enhancer prediction, three datasets are provided for identifying enhancer regions in the mouse and human genomes. Regarding species classification, two datasets are selected to identify sequences as either coding (exonic) or intergenic (non-coding) and to classify sequences as originating from humans or worms (C. elegans). Regarding the classification of regulatory elements, three datasets are utilized to categorize sequences as regulatory regions based on Ensembl annotations, identify open chromatin regions, or pinpoint non-TATA promoter regions in the human genome. Each example is a 200bp central sequence extracted from the reference genome and labeled by high-throughput functional assays.
We utilize the fully reproduced results of various DNA models in GenBench \citep{liu2024genbench}. Table~\ref{tab:app_genomic_benchmark} provides full results of the Genomics Benchmark.

\subsection{DNA Tasks with Nucleotide Transformer (NT) Benchmark}
The NT benchmark~\citep{dalla2023nucleotide} spans 18 diverse tasks that probe both local and distal regulatory logic, including ten histone-mark predictions (\textit{e.g.}, H3K4me3), four promoter/enhancer subtasks, and four canonical splice-site detection variants. We adopt the original version of the NT benchmark, where sequences of 1–4~kbp surrounding epigenetic peaks are sampled from the hg38 assembly with chromosome-wise train/valid/test splits to prevent leakage. We use \textbf{Matthews Correlation Coefficient (MCC)} for imbalanced or noisy labels (histone marks, splice sites) and \textbf{F1 score} for balanced promoter/enhancer tasks. Table~\ref{tab:nt_benchmark} presents the full results of the NT benchmark with popular DNA models.

\subsection{DNA Tasks with GUE Benchmark}
As proposed by DNABERT2~\citep{iclr2024dnabert2}, the GUE benchmark contains 24 datasets of 7 practical biological genome analysis tasks for 4 different species.
All sequences are provided as FASTA strings (or BED coordinates) with an official 70/15/15 train/valid/test split; inputs range from 70 bp to 1 kb so that both local and distal regulatory logic is probed.  We follow the original evaluation protocol and report MCC for all binary tasks, while the multi-class Covid variant classifier is scored with macro-F1. A concise task-wise description is given below.

\textbf{(A) Promoter Detection (Human).}
Identify proximal promoters ($-$249~+50~bp around TSS) in the GRCh38 assembly.  Positive examples come from EPDnew~\citep{dreos2013epdpromoter}; negatives are random intergenic regions matched by GC content, stratified into \emph{TATA}, \emph{non-TATA}, and their union (\emph{all})—yielding \textbf{3 datasets}.
\textbf{(B) Core-Promoter Detection (Human).}
Predict the 70~bp core promoter window ($-$34~+35~bp) immediately flanking the TSS, a harder variant of (A) due to shorter context.  Positive/negative selection mirrors (A), giving \textbf{3 additional datasets}.
\textbf{(C) Transcription-Factor Binding Site Prediction (Human).}
Classify 101~bp regions centered on ChIP-seq peaks from ENCODE (161 TFs, 91 cell lines).  GUE retains \textbf{5 representative TFs} after filtering trivial or extremely imbalanced cases; negatives are GC-matched genomic windows outside peaks.
\textbf{(D) Splice-Site Prediction (Human).}
Detect splice \emph{donor} and \emph{acceptor} sites within 400~bp windows extracted from GRCh38.  To avoid saturation, the original 10k-sample dataset~\citep{BMC2021spliceator} is adversarially augmented with hard negatives until MCC stabilizes, producing \textbf{1 challenging dataset}.
\textbf{(E) Transcription-Factor Binding Site Prediction (Mouse).}
Analogous to (C) but using mouse ENCODE ChIP-seq (78 TFs).  Five TFs are randomly chosen following the same GC-matched negative selection, resulting in \textbf{5 datasets}.
\textbf{(F) Epigenetic-Mark Prediction (Yeast).}
Predict presence of 10 histone marks or nucleosome occupancy tracks in \emph{S.\ cerevisiae}.  Each sequence is 147~bp centered on experimentally validated sites downloaded from the JAIST repository; random genomic positions serve as negatives, yielding \textbf{10 datasets}.
\textbf{(G) Covid Variant Classification (Virus).}
Multi-class identification of \emph{SARS-CoV-2} lineages (\textit{Alpha}, \textit{Beta}, ..., \textit{Zeta}) from 1~kb genome snippets obtained via GISAID’s EpiCoV~\citep{khare202SARS_CoV_2}.

\subsection{Multi-omics Downstream Tasks}
\label{app:multi_omics}
\paragraph{RNA Splicing Site Prediction}
SpliceAI~\citep{JAGANATHAN2019SpliceAI} pairs 10kb sequences centered on annotated splice junctions with binary labels (\textit{donor}, \textit{acceptor}).  
Following \cite{liu2024genbench}, we use the ``chromosome-held-out" split: chromosomes 1–19 for training, 20 for validation, and 21–22 for testing, guaranteeing isoform-level independence. Performance is evaluated by the Area Under the ROC Curve (AUROC) for each type of site and averaged between donor/acceptor.

\paragraph{Long-Range Benchmarks (LRB)}
We adopt two long-range tasks collected in LRB~\cite{trop2024LRB}.
\textbf{(a) Causal eQTL Variant Effect.} Given a 20kb locus and a candidate SNV, the task is to classify whether the variant modulates gene expression; ground truth is derived from GTEx v8 fine-mapping.  We report AUROC on the held-out tissue set proposed by \cite{trop2024LRB}.  
\textbf{(b) Bulk RNA Expression.} Predict log-transformed gene expression in 54 GTEx tissues from a 40~kb upstream sequence window.  Following LRB, we fit a linear regressor on frozen sequence embeddings and compute the coefficient of determination ($R^2$) on an unseen chromosome split.

\paragraph{Zero-shot Protein Fitness Prediction}
Deep Mutational Scanning (DMS)~\cite{alquraishi2019proteinnet} assays exhaustively mutate a protein coding sequence and measure functional fitness. We follow Evo~\cite{science2024EVO} and evaluate two representative datasets: Escherichia coli TEM-1 $\beta$-lactamase (Bacteria) and human BRCA1 RING domain (Human). All single- and double-mutant variants are used. True fitness values remain hidden during inference to emulate zero-shot generalization. We report \textbf{Spearman’s Rank Correlation Coefficient (SRCC)} between model scores (negative log-likelihoods of mutated codons) and experimental fitness.

\section{Extended Related Work}
\label{app:extend_related}
\subsection{DNA Foundation Models}
Researchers have started to build large-scale sequence models for genomes since the 2020s, inspired by advances in natural language processing~\cite{vaswani2017attention, Touvron2023LLaMA}. Early attempts like DNABERT~\cite{ji2021dnabert} and DNAGPT~\cite{zhang2023dnagpt} demonstrated that Transformer architectures can be adapted to DNA by treating nucleotide sequences as a ``language". Subsequent models greatly scaled up pre-training (GPN~\cite{benegas2023GPN}, Nucleotide Transformer variants~\cite{dalla2023nucleotide}, and DNABERT2~\cite{iclr2024dnabert2}), achieved genome-scale pre-training on human or multi-species data, and showed broad utility across diverse downstream genomic tasks~\cite{BMC2023genomicbenchmark, 2025DNALongBench}. The methodologies of these DNA language models can be discussed along four key dimensions: (a) \textit{long sequence modeling}, (b) \textit{DNA tokenization strategies}, (c) \textit{pre-training objectives}, and (d) \textit{pre-training domains and downstream tasks}.

\paragraph{Long Sequence Modeling.}
A core technical challenge is modeling extremely long sequences (\textit{e.g.}, many gene regions span 8k–1M bases). Several works have adopted state-space models (SSMs)~\cite{Gu2023Mamba, yang2024gateddelta} for efficient long-range modeling: HyenaDNA~\cite{nguyen2024hyenadna}, MSAMamba~\cite{thoutam2024MSAMamba}, and Caduceus~\cite{schiff2024caduceus} were among the first to use linear-time SSM architectures for DNA, enabling context lengths beyond what vanilla Transformers~\cite{Dao2022FlashAttention} can handle. More recently, MegaDNA~\cite{NC2024MegaDNA} introduced a hierarchical Transformer for genome sequences, while models like Evo2~\cite{broxiv2025EVO2}, LifeCode~\cite{Liu2025LifeCode}, and HybriDNA (2025)~\cite{Ma2025HybriDNA} combined SSMs with self-attention mechanisms to balance memory efficiency and modeling accuracy~\cite{yang2024gateddelta}. Meanwhile, alternative sequence learners have been tried: ConvNova~\cite{iclr2025convnova} applies convolutional neural networks to genomic sequences, and DeepGene~\cite{zhang2024deepgene} leverages a graph neural network over a pan-genome graph to encode DNA context.

\paragraph{DNA Vocabulary and Tokenization.}
Unlike natural language, DNA has no inherent word boundaries or semantics, and coding \textit{vs.} non-coding regions have different significance (coding regions translate in triplets as codons)~\cite{cooper1981centraldogma, nmil2024CaLM}. Accordingly, most recent DNA models~\cite{science2024EVO, nguyen2024hyenadna} forego complex tokenization and simply use the four nucleotides (A, C, G, T) as the base vocabulary. Some approaches employ fixed-length k-mers or subwords: DNABERT~\cite{ji2021dnabert} and Nucleotide Transformer~\cite{dalla2023nucleotide} represent sequences as k-mer tokens, and DNABERT2~\cite{iclr2024dnabert2} and GENA-LM~\cite{ji2023genalm} build a byte-pair encoding (BPE) vocabulary for DNA. Beyond static schemes, researchers have proposed dynamic data-driven tokenization, \textit{e.g.}, VQDNA~\cite{icml2024VQDNA} and MxDNA~\cite{nips2024MXDNA} learn custom DNA tokens via Vector Quantization (VQ)~\cite{2017VQ-VAE, iclr2024MAPEPPI, iclr2025MeToken} and deformable convolution~\cite{iccv2017Deformable}, automatically discovering higher-level nucleotide motifs.

\paragraph{Pre-training Objectives.}
The training strategies for DNA foundation models mirror those in NLP~\cite{devlin2019bert}, mainly falling into masked language modeling (BERT-style) or autoregressive generation~\cite{Radford2018GPT1}. Using nucleotide k-mers or subwords, encoder-based models like DNABERT2 and GENERanno~\cite{li2025generanno} employ masked sequence modeling (predicting masked bases or k-mers from context), which has proven effective for many short-range genomic annotation tasks~\cite{iclr2024dnabert2}. In contrast, many large parameter DNA models adopt autoregressive training to better model long-range dependencies~\cite{ji2023genalm, benegas2023GPN}. Evo2~\cite{broxiv2025EVO2} further adjusts loss weights between coding and non-coding regions to emphasize functionally important context. Meanwhile, GeneMask~\cite{ECAI2023GeneMask} and CM-GEMS~\cite{ECAI2024CMGEMS} introduce adaptive or focused masking schemes to accelerate pre-training and improve sequence representation. Beyond reconstruction-based objectives, some works pursue alternative pre-training tasks: DNABERT-S~\cite{zhou2025dnaberts} uses contrastive learning~\cite{chen2020simclr} to learn species-aware embeddings, and LifeCode~\cite{Liu2025LifeCode} incorporates knowledge distillation and cross-modal alignment into pre-training to fuse information across the central dogma (DNA–RNA–protein).

\paragraph{Pre-training Domains and Downstream tasks.}
The choice of pre-training data strongly influences a model’s generalization ability. Most DNA foundation models have been trained on the human genome~\cite{nguyen2024hyenadna} or a mixture of species~\cite{iclr2024dnabert2}, enabling them to generalize across common genomic tasks in eukaryotes. However, some models target specific taxonomic domains: Evo~\cite{science2024EVO}, for example, was trained on prokaryotic genomes and excels at bacterial and viral tasks such as designing CRISPR–Cas9 target sequences. Likewise, AgroNT~\cite{mendoza2023AgroNT}, PlantCaduceus~\cite{zhai2025PlantCaduceus}, and PDLLMs~\cite{liu2025pdllms} are specialized on plant genomic data, and ProKaformer~\cite{AAAI2025ProKaFormer} is tailored to model microbial community (microbiome) genomes. DNABERT-S~\cite{zhou2025dnaberts} builds species-differentiated embeddings for multi-species data, and GenomeOcean~\cite{zhou2025genomeocean} leverages large-scale metagenomic assemblies to learn generalizable genome representations.
After pre-training, DNA foundation models are evaluated on a range of downstream tasks. Common tasks include token-level predictions and sequence-level classifications or regressions (\textit{e.g.}, predicting regulatory function or phenotype from a sequence). For instance, SegmentNT~\cite{de2024segmentnt} uses a foundation model to partition genomes into meaningful segments at single-nucleotide resolution. Meanwhile, GPT models~\cite{science2024EVO, zhu2024CDGPT} can generate realistic DNA sequences (even whole genomes) in an autoregressive manner, and diffusion-based approaches \cite{Li2024DiscDiff} produce novel DNA sequences via latent diffusion. Finally, researchers are exploring hybrid AI systems that integrate general-purpose LLMs~\cite{Bai2023Qwen} with genomic expertise~\cite{Liu2024GenoTEX, li2025DNAOmni} and large parameter scales~\cite{aaai2026trinitydna}, which combine open-source ChatGPT-like models with DNA foundation models to handle complex multi-step genomic applications.

\subsection{Multi-omics Modeling}
Beyond DNA-alone modeling, researchers are extending foundation models to multi-omics~\cite{krishna2024RFdiffusionAA, he2024lucaone}, aiming to connect DNA with other molecular modalities like RNA and protein within the unified framework~\cite{hayes2025ESM3, science2024EVO}. Protein-centric models have already achieved remarkable success in structure prediction and design (\textit{e.g.}, AlphaFold-2~\cite{Nat2021AlphaFold}), which motivates developing DNA-centric multi-omics models that leverage genomic information. Recent research~\cite{ji2023genalm, yang2024CREformer} uses DNA sequence models to predict downstream molecular phenotypes (\textit{e.g.}, protein function or chromatin profiles) across species. Based on the central dogma principle~\cite{cooper1981centraldogma}, current DNA-centered multi-omics models generally fall into two broad categories.

The \textit{first category} seeks to predict protein-level properties and functions directly from DNA sequences, effectively learning the end-to-end mapping from genome to proteome. Evo~\cite{science2024EVO} pioneered this approach by modeling sequence-to-function relationships at the genome scale, and its successor Evo2~\cite{broxiv2025EVO2} further expanded to modeling genomes across all domains of life. AIDO~\cite{Song2024AIDO} and HybriDNA~\cite{Ma2025HybriDNA} are designed to handle ultra-long inputs (full genomes), CD-GPT~\cite{zhu2024CDGPT}, LifeCode~\cite{Liu2025LifeCode}, and GENERator~\cite{Wu2025GENERator} incorporate explicit transcription and translation tasks into their objectives, and frameworks like LucaOne~\cite{he2024lucaone} use supervised multi-task learning to unify representations of nucleic acids and proteins for diverse predictive tasks. They could be further extended to Metagenomic applications with multi-omics data~\cite{zhou2025dnaberts, BiB2025FGBERT}.
The \textit{second category} allows gene-to-expression modeling, focusing on predicting cellular and molecular readouts from DNA sequence~\cite{nm2021enformer, avsec2025alphagenome} and bridging genotype to phenotype at the regulatory or cellular level. Enformer~\cite{nm2021enformer} first demonstrated that a deep Transformer can accurately predict gene expression profiles from DNA by accounting for long-range genomic interactions. Geneformer~\cite{nature2023geneformer} showed the benefit of transfer learning for network biology predictions, IsoFormer~\cite{nips2024IsoFormer} and SPACE~\cite{icml2025space} introduced multi-modal alignment and mixture-of-experts to enhance regulatory sequence predictions, and AlphaGenome~\cite{avsec2025alphagenome} and CDBridge~\cite{2025cdbridge} were recently proposed as a unified model that can predict a wide range of cell-level genomic assay results from the long DNA context.

\subsection{Byte Architectures}
Most traditional language models rely on a pre-computed tokenizer to convert raw data into tokens—commonly using subword algorithms like BPE~\cite{Sennrich2015BPE} or SentencePiece~\cite{EMNLP2018SentencePiece} to segment the input byte sequence. Complex dynamic tokenization methods, such as Dynamic Pooling~\cite{ACL2023DynamicPool} and Dynamic Vocabulary~\cite{EMNLP2024DynVoc}, still involve training separate tokenization modules and do not fundamentally escape the constraints of discrete token boundaries. This tokenization step might introduce information loss or segmentation artifacts, motivating a shift toward models that operate directly on byte sequences~\cite{Pagnoni2024BLT, hwang2025hnet}.
With the advent of more efficient long-context sequence models~\cite{Gu2023Mamba, yang2025deltanet, ijcai2024longvq}, a new class of byte-level architectures has emerged to eliminate tokenization altogether~\cite{COLM2024Mambabyte}. MegaByte~\cite{NIPS2023MEGABYTE} and SpaceByte~\cite{NIPS2024SpaceByte} pioneered multiscale Transformers and SSMs that process sequences at the byte level, enabling context lengths on the order of millions of characters without arbitrary tokenization. Building on this, ByteScale~\cite{Ge2025ByteScale} scaled up token-free models to unprecedented sizes (allowing training with 2-million token contexts on thousands of GPUs), while bGPT~\cite{wu2024bGPT} extended the tokenizer-free approach to multimodal data. Very recently, BLT~\cite{Pagnoni2024BLT} performs entropy-based segmentation of byte sequences using a latent Perceiver-style architecture~\cite{icml2021Perceiver, iclr2022PerceiverIO}, whereas HNet~\cite{hwang2025hnet} provides a fully differentiable tokenization scheme, enabling end-to-end learning of how to chunk byte sequences during model training.

\end{document}